\documentclass[prd,aps,preprintnumbers,nofootinbib,twocolumn]{revtex4}
\usepackage{epsfig}
\usepackage{amsmath}
\usepackage{hyperref}
\usepackage{physics}
\usepackage{xcolor}
\usepackage{accents}

\def\as{\alpha_S}

\begin{document}

\preprint{Cavendish-HEP-21/09, P3H-21-043, TTK-21-20}

\renewcommand{\thefigure}{\arabic{figure}}

\title{Tour de force in Quantum Chromodynamics:\\ A first next-to-next-to-leading order study of three-jet production at the LHC}

\author{Micha\l{}  Czakon}
\affiliation{{\small Institut f\"ur Theoretische Teilchenphysik und Kosmologie, RWTH Aachen University, D-52056 Aachen, Germany}}
\author{Alexander Mitov}
\affiliation{{\small Cavendish Laboratory, University of Cambridge, Cambridge CB3 0HE, UK}}
\author{Rene Poncelet}
\affiliation{{\small Cavendish Laboratory, University of Cambridge, Cambridge CB3 0HE, UK}}

\date{\today}

\begin{abstract}
Multi-jet rates at hadron colliders provide a unique possibility for probing
Quantum Chromodynamics (QCD), the theory of strong interactions. By comparing
theory predictions with collider data, one can directly test perturbative QCD,
extract fundamental parameters like the strong coupling $\alpha_s$ and search
for physics beyond the Standard Model. In this work we calculate, for the first
time, the next-to-next-to-leading (NNLO) QCD corrections to typical three-jet
observables and to differential three-to-two jet ratios. The calculation is
complete apart from the three-jet double virtual contributions which are
included in the leading-colour approximation. We demonstrate that the inclusion
of the NNLO corrections significantly reduces the dependence of those
observables on the factorization and renormalization scales.  Besides its
phenomenological value, this proof-of-principle computation represents a
milestone in perturbative QCD.
\end{abstract}

\maketitle

\section{Introduction}\label{sec:introduction}
The production of highly energetic sprays of particles, also known as jets, is
a dominant process at hadron colliders. At high energies, where perturbation
theory is expected to hold, this process offers the possibility for studying
QCD in great detail. The theory--data comparison of differential multi-jet
rates provides essential information about perturbative QCD and the modeling of
jet production. The precision of these predictions is typically limited by
their dependence on unphysical parameters -- such as the renormalization and
factorization scales -- but it can be systematically increased by including
higher-order corrections.

Three-jet production at the Large Hadron Collider (LHC) has been studied in
great detail by experimental collaborations, see for example
ref.~\cite{Aad:2011tqa, Aad:2014rma, CMS:2014mna,
ATLAS:2015yaa,Aaboud:2018hie,Aad:2020fch}. Typical observables are jet
transverse momenta, angular correlations and, more generally, event-shape
observables. A particular type of observable suited for perturbative QCD is the
ratio $R_{3/2}$ of three-to-two jet rates \cite{Chatrchyan:2013txa}.  These
ratios are directly sensitive to parton splittings and are, therefore,
proportional to the strong coupling constant $\alpha_s$. This provides an
opportunity for measuring the value of $\alpha_s$ at the LHC. Cross section
ratios have the additional advantage that some systematic uncertainties of
experimental and theoretical nature cancel out. A prime example is the
dependence on parton distribution functions (pdf).

There is extensive literature on theoretical predictions for multi-jet
production through NLO in perturbative QCD \cite{Ellis:1992en,
Giele:1993dj,Nagy:2001fj, Nagy:2003tz, Bern:2011ep, Badger:2013yda}, including
NLO electroweak corrections \cite{Dittmaier:2012kx,Frederix:2016ost,Reyer:2019obz}. NLO
computations have also been matched to parton-showers
\cite{Alioli:2010xa,Hoeche:2012fm} and are generally present in multi-purpose
event generators \cite{Gleisberg:2008ta, Alwall:2014hca, Frederix:2018nkq}.
Higher-order predictions for two-jet and single-inclusive jet production have
seen extensive development in the past decade and are implemented through NNLO
in QCD \cite{Currie:2016bfm, Currie:2017eqf, Czakon:2019tmo,
AbdulKhalek:2020jut}. The feasibility of NNLO QCD predictions for higher jet
multiplicity is limited by the availability of two-loop virtual amplitudes and
by the efficient treatment of real radiation contributions. The three-jet
two-loop amplitudes have recently been made public in the leading-color
approximation \cite{Chicherin:2020oor, Abreu:2021fuk}, leaving the real
radiation as the last obstacle to predictions accurate at second order in $\as$.

The aim of this article is twofold. Firstly, it presents NNLO QCD predictions
for the production of three jets and $R_{3/2}$ ratios at the LHC at 13 TeV.
Secondly, it demonstrates the technical ability to treat the NNLO real
radiation contributions for processes with five colored partons at the Born
level.  The completion of the second order corrections to three jet production
is a milestone in perturbative QCD computations since, judging by its infra-red
structure, it is among the most complicated two-to-three processes at the LHC.

This paper is organized as follows: in section~\ref{sec:calculation} we discuss
the technical details of our computation. Section~\ref{sec:results} contains
the phenomenological results and their analysis.  We conclude with a summary
and outlook on future applications in section~\ref{sec:conclusion}.

\section{Calculation details}\label{sec:calculation}
The computation has been performed within the sector-improved residue
subtraction scheme formalism \cite{Czakon:2010td,Czakon:2014oma} which has
already been successfully applied to single inclusive jet production
\cite{Czakon:2019tmo} and many other hadron collider processes, see
refs.~\cite{Czakon:2015owf, Chawdhry:2019bji, Chawdhry:2021hkp}. We work in
five-flavour massless QCD without the top quark. Tree-level matrix elements
have been taken from the {\tt AvH} library \cite{avhlib,Bury:2015dla} while all
one-loop matrix elements have been implemented with the {\tt OpenLoops} library
\cite{Buccioni:2019sur}. The double virtual matrix elements are not yet
available beyond the leading-colour approximation. For this reason we
approximate the finite two-loop contribution
\begin{eqnarray}
 \mathcal{R}^{(2)} (\mu_R^2) &=&
   2 \Re\left[ \mathcal{M}^{\dagger (0)}\mathcal{F}^{(2)}\right] (\mu_R^2)
   + \big\vert \mathcal{F}^{(1)}\big\vert^2(\mu_R^2)\nonumber\\
   &\equiv & {\cal R}^{(2)}(s_{12})+\sum_{i=1}^4 c_i
                           \ln^i\left({\mu_R^2\over s_{12}}\right)\,,
\end{eqnarray}
where $s_{12} = (p_1+p_2)^2$ the invariant mass of the incoming partons,
in the following way
\begin{equation}
{\cal R}^{(2)}(s_{12}) \approx {\cal R}^{(2) l.c.}(s_{12})\,,
\label{eq:approx}
\end{equation}
where ${\cal R}^{(2) l.c.}(s_{12})$ denotes its leading-colour approximation.
It is taken from the {\tt C++} implementation provided in
ref.~\cite{Abreu:2021fuk}.

Eq.~(\ref{eq:approx}) above is the only approximation made in the present
computation.  We have checked that the overall contribution of ${\cal R}^{(2)
l.c.}(s_{12})$ is about $10\%$ and we expect the missing pure virtual
contributions beyond the leading-colour approximations to be further
suppressed.

We consider production of two and three jets at the LHC with a center of mass
energy of 13 TeV with jet requirements adapted from experimental phase space
definitions like, for example, ref.~\cite{Aad:2020fch}. Jets are clustered
using the anti-$k_T$ algorithm \cite{Cacciari:2008gp} with a radius of $R =
0.4$ and required to have transverse momentum $p_T(j)$ of at least $60$ GeV and
rapidity $y(j)$ fulfilling $|y(j)| < 4.4$. All jets passing this requirement
are sorted and labeled according to their $p_T$ from largest to smallest.
Among those jets we require the two leading jets to fulfill $p_T(j_1) +
p_T(j_2) > 250$ GeV in order to avoid large higher-order corrections in two-jet
production close to the phase space boundary.  We denote by $\dd\sigma_n$ the
differential cross section for at least $n$ jets fulfilling the above criteria.
Its expansion in $\as$ reads
\begin{align}
  &\dd\sigma_n =  \dd\sigma_n^{(0)}+\dd\sigma_n^{(1)}+\dd\sigma_n^{(2)}
              +\order{\as^{n+3}}\;\nonumber\\
  &\dd\sigma_n^{\rm LO} = \dd\sigma_n^{(0)}\;,\nonumber\\
  &\dd\sigma_n^{\rm NLO} = \dd\sigma_n^{(0)}+\dd\sigma_n^{(1)}\;,\nonumber\\
  &\dd\sigma_n^{\rm NNLO} = \dd\sigma_n^{(0)}+\dd\sigma_n^{(1)}+\dd\sigma_n^{(2)}\;.
\end{align}

We quantify the size of (N)NLO corrections with the help of the following
ratios of differential cross sections
\begin{align}
 K^{\rm NNLO} = \frac{\dd\sigma^{\rm NNLO}}{\dd\sigma^{\rm NLO}} \;
   \quad \text{and} \quad
 K^{\rm NLO} = \frac{\dd\sigma^{\rm NLO}}{\dd\sigma^{\rm LO}}\;.
\end{align}

The pdf set {\tt NNPDF31\_nnlo\_as\_0118} is used for all perturbative orders.
The renormalization $\mu_R$ and factorization $\mu_F$ scales are set equal
$\mu_R = \mu_F = \mu_0$. The central scale $\mu_0$ is chosen as $\hat{H}_T/n$
for $n=1,2$, where
\begin{equation}
  \hat{H}_T = \sum_{i \in \text{partons}} p_{T,i} \,.
\end{equation}
The sum in the above equation is over all final state partons, irrespective of
the jet requirements. Previous studies of perturbative convergence in jet
production support this event-based dynamic scale \cite{Currie:2018xkj,
Bellm:2019yyh}. Unless stated otherwise, uncertainties from missing higher
orders in perturbation theory are estimated by variation of $\mu_F = \mu_R$
by a factor of $2$ around the central scale $\mu_0$.

The calculation of the three-jet production cross sections is computationally
and technically challenging. The main bottlenecks are the double real radiation
corrections and the corresponding integrated subtraction terms, due to large
numerical cancellation between individual contributions. The numerical
evaluation of the complex double-virtual amplitudes is fast due to the efficient
representation presented in \cite{Abreu:2021fuk}.

\section{Results}\label{sec:results}

\begin{figure}[t]
  \centering
  \includegraphics[width=0.5\textwidth]{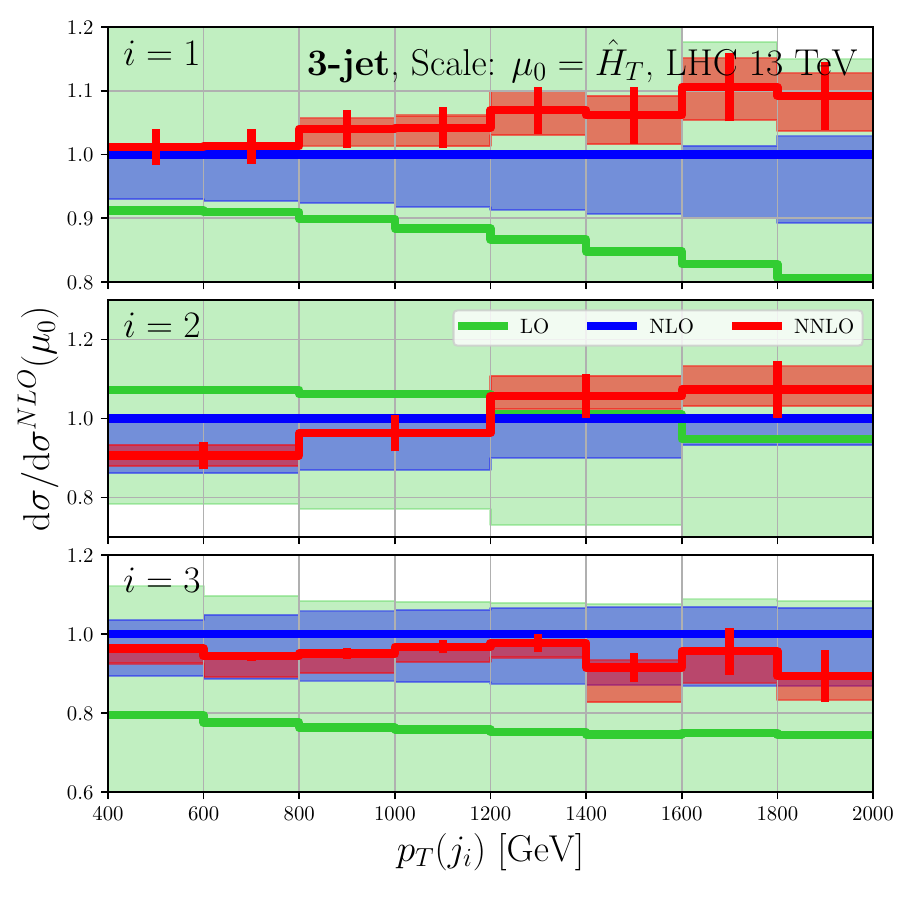}
  \caption{The three panels show the $i$th leading jet transverse
           momentum $p_T(j_i)$ for $i=1,2,3$ for the production of (at least)
           three jets. LO (green), NLO (blue) and NNLO (red) are shown
           for the central scale (solid line). 3-point scale variation is
           shown as a coloured band. The grey band corresponds to the
           uncertainty from Monte Carlo integration.}
  \label{fig:pTi}
\end{figure}

We begin by discussing typical jet observables at hadron colliders. In
fig.~\ref{fig:pTi} we show differential cross sections for three-jet production
with respect to the transverse momentum $p_T(j_i)$ of the $i$th leading jet.
In all histograms the outer bins do not include over- or under-flow events.

The NNLO $K$-factor of the $p_T(j_1)$ distribution is not flat: at small
$p_T(j_1)$ one observes small positive NNLO corrections, while at large
$p_T(j_1)$ the corrections tend to be about $ 10\%$ and positive. The
change in scale dependence for this observable when going from NLO to NNLO is
also dependent on $p_T(j_1)$. One observes a small reduction at large
$p_T(j_1)$ (from about $7\%$ at NLO to about $5\%$ at NNLO) while at small
$p_T(j_1)$, where the $K$-factor is smallest, the scale dependence decreases
significantly (from about $4\%$ at NLO to about $1\%$ at NNLO). Interestingly,
the scale dependence at NLO and NNLO behaves similarily: it steadily increases
with $p_T(j_1)$.  Throughout this work we define the scale dependence as one
half of the width of the scale uncertainty band. This is relevant for cases
where the scale variation is asymmetric, as for example is the case of
$p_T(j_1)$ at NLO.

The $p_T(j_2)$ distribution has a different pattern of NNLO corrections: relative
to NLO they are negative at low $p_T(j_2)$, and steadily
increase towards larger $p_T(j_2)$ values. At both NLO and NNLO the scale
dependence of $p_T(j_2)$ is similar to that of $p_T(j_1)$.  On the technical
side, the convergence of the numerical integration for the $p_T(j_2)$ spectrum
has been significantly slower than for the other $p_T$ observables, which
results in increased Monte Carlo uncertainty. To compensate for this, a larger
bin size has been used for $p_T(j_2)$. Independently of its slower numerical
convergence, the $p_T(j_2)$ spectrum shows good perturbative convergence. Such
a behavior is in contrast to the two-jet case where the sub-leading $p_T$
spectrum is known to get large perturbative corrections due to the strict
back-to-back tree-level kinematics \cite{Currie:2018xkj}.

The $p_T(j_3)$ distribution is well-behaved: it has a flat $K$-factor and
fairly symmetric uncertainty band at both NLO and NNLO. The scale variation is
almost independent of $p_T(j_3)$, about the $5\%$ at NNLO, which is
significantly smaller than in the NLO case.

The fact that this observable shows such good convergence and perturbative
stability is somewhat remarkable. Naively, one may suspect that the scale used
here may not perform very well for this distribution because the scale is based
on the kinematics of the full event which, in turn, is dominated by the leading
jet(s). Since this finding may be of relevance for the extraction of the strong
coupling constant from three-jet events, it may be worth investigating this
behavior in more detail. This is outside the scope of the present work.

\begin{figure}[t]
  \centering
  \includegraphics[width=0.5\textwidth]{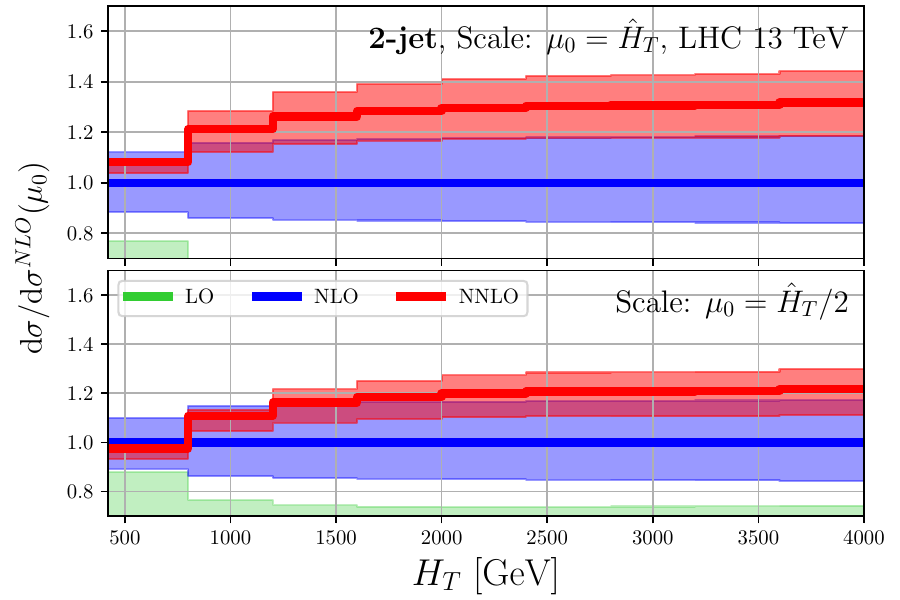}
  \caption{The observable $H_T$ in two-jet
           production for two different
           central scale choices. Scale variation corresponds
           to 3-point variation. The colours are the same as in
           fig.~\ref{fig:pTi}.}
  \label{fig:HT_2j}
\end{figure}

\begin{figure}[t]
  \centering
  \includegraphics[width=0.5\textwidth]{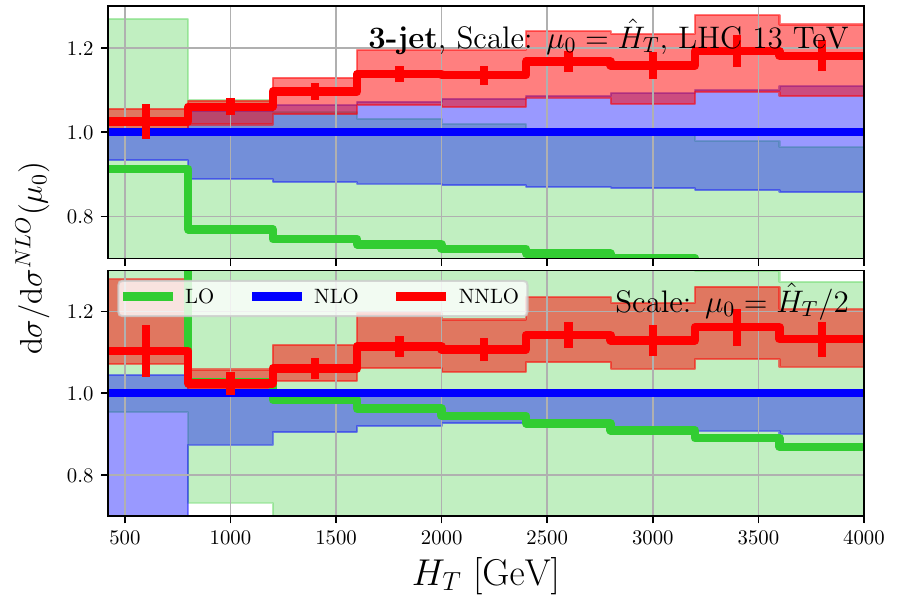}
  \caption{As in fig.~\ref{fig:HT_2j} but for three-jet production.}
  \label{fig:HT_3j}
\end{figure}

Next we discuss the observable $H_T$, defined as
\begin{equation}
  H_T = \sum_{i \in \text{jets}} p_T(j_i)\;,
\end{equation}
where the sum is over all jets that pass the jet requirements. We show this
observable in fig.~\ref{fig:HT_2j} for the two-jet process and in
fig.~\ref{fig:HT_3j} for the three-jet process. Both figures are subdivided in
two panels showing the same observable but for a different central scale
choice: the upper panels for $\mu_0 = \hat{H}_T$ and the lower panels for
$\mu_0 = \hat{H}_T/2$. Turning to the two-jet case we see
that both the perturbative convergence and the scale dependence improve if the
central scale choice is lowered. For $\mu_0 = \hat{H}_T$ the inclusion of the
NNLO QCD corrections does not reduce significantly the scale dependence with
respect to NLO and both bands barely overlap.  However, $K^{\rm NNLO} \approx
1.2$ is much smaller than $K^{\rm NLO} \approx 2$ indicating the stabilization
of higher-order corrections beyond NNLO.  For the production of three jets we
find that the two central scale choices $\mu_0 = \hat{H}_T$ and $\mu_0 =
\hat{H}_T/2$ produce comparable results, albeit $\mu_0 = \hat{H}_T/2$ leads
to NNLO QCD corrections which are not captured by the NLO scale band.
The scale dependence and $K^{\rm NNLO}$ is similar compared to the two-jet
case. These findings indicate that a
central scale of $\mu_0 = \hat{H}$ leads to slightly better perturbative
convergence and, thus, better approximates the actual energy scale relevant for
this observable. We have checked that even lower scales, like $\mu_0 =
\hat{H}_T/4$, spoil perturbative convergence.

As a first application of the NNLO--accurate three-jet rates computed in this
work we consider ratios between three-jet and two-jet rates. The ratios are
defined as
\begin{equation}
 R_{3/2}(X,\mu_R,\mu_F) = \frac{\dd\sigma_3(\mu_R,\mu_F)/\dd X}
                               {\dd\sigma_2(\mu_R,\mu_F)/\dd X}\;,
\end{equation}
where $X$ is some observable of interest. The (N)NLO ratio is defined in such a
way that the numerator and denominator on the right hand side are evaluated at
the matching order. The scale dependence of the differential cross sections is
shown explicitly to emphasize that the scale choices in the numerator and
denominator are correlated.

\begin{figure}[t]
  \centering
  \includegraphics[width=0.5\textwidth]{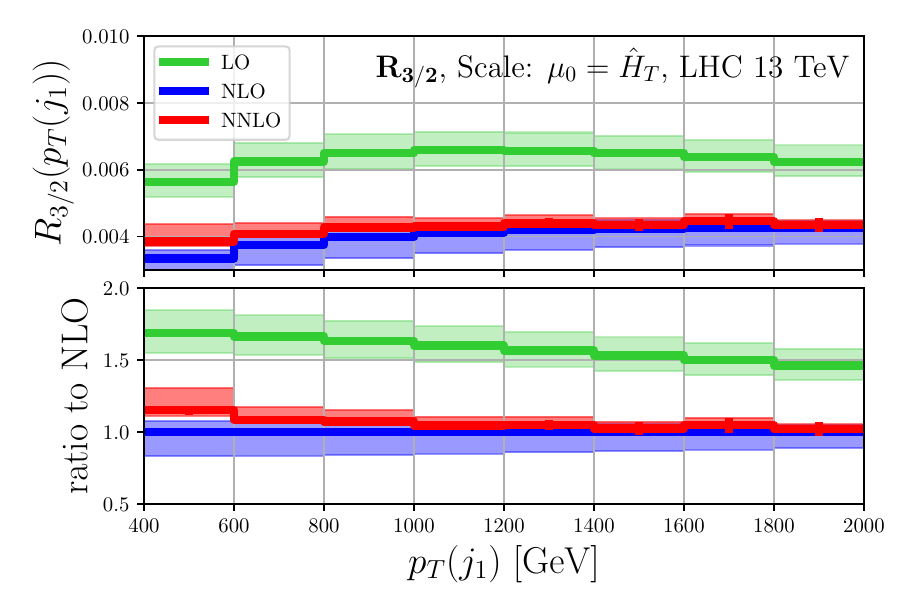}
  \includegraphics[width=0.5\textwidth]{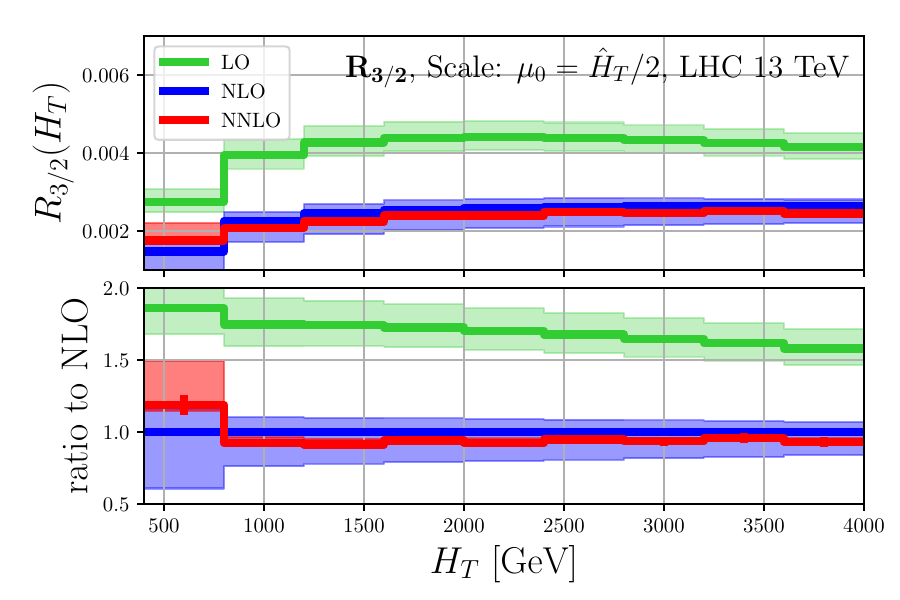}
  \caption{The top two panels show $R_{3/2}(p_T(j_1))$ (in absolute and
           as ratio to NLO) and the bottom two panels $R_{3/2}(H_T)$.
           The colours are the same as in fig.~\ref{fig:pTi}.}
  \label{fig:r32_pT_and_HT}
\end{figure}

In the upper two panels of fig.~\ref{fig:r32_pT_and_HT} we show the ratio
$R_{3/2}(p_T(j_1))$. The ratio changes drastically when going from LO to NLO
mostly due to the change in the two-jet cross section. The NNLO correction
stabilizes the ratio and leads to a very small scale dependence.  The $K^{\rm
NNLO}$ factor slightly decreases for large momenta, however, it is always fully
contained within the NLO scale band. An important observation is that the NNLO
scale band is very small in comparison to NLO, reducing it from about $10\%$
down to $3\%$.

Next we consider the lower two panels in fig.~\ref{fig:r32_pT_and_HT}, where we
show the ratio $R_{3/2}(H_T)$ for a central scale $\mu_0 = H_T/2$.  This
observable behaves similarly to $R_{3/2}(p_T(j_1))$ albeit with a slightly
larger scale dependence.  The reduction in the scale uncertainty when going
from NLO to NNLO is of particular importance since this observable is used
experimentally for measurements of $\as$ \cite{Aaboud:2018hie}. The leading
source of perturbative uncertainty in this data--theory comparison is the scale
dependence. The pdf dependence, which is not computed in this work, is expected
to be reduced in the ratio.

Jet rates are typically measured in slices of jet rapidity. To demonstrate how
our calculation performs in this situation, we divide the phase space in slices
of the rapidity difference between the two leading jets
\begin{equation}
  y^* = |y(j_1) - y(j_2)|/2\;,
\end{equation}
and define the ratio of the two- and three-jet rates as
\begin{equation}
 R_{3/2}(H_T,y^*) = \frac{\dd^2\sigma_3/\dd H_T/\dd y^*}
                         {\dd^2\sigma_2/\dd H_T/\dd y^*}\;.
\label{eq:r32_ht_ys}
\end{equation}

\begin{figure}[t]
  \centering
  \includegraphics[width=0.5\textwidth]{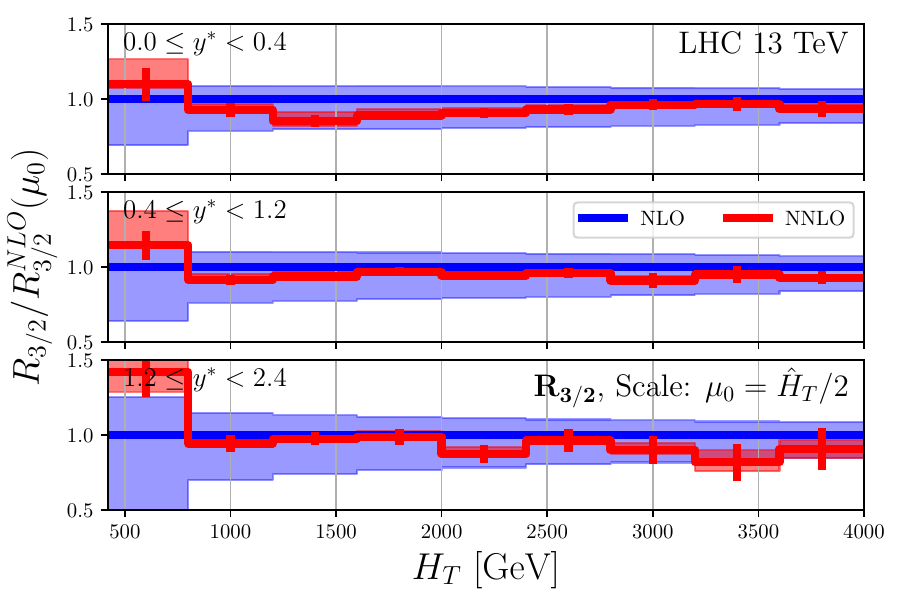}
  \caption{The three panels show $R_{3/2}(H_T,y^*)$, in each
           panel a different slice in $y^*$ as ratio to NLO.
           The colours are the same as in fig.~\ref{fig:pTi}.}
  \label{fig:r32_ystarxHT}
\end{figure}

The NNLO prediction for this cross section ratio can be found in
fig.~\ref{fig:r32_ystarxHT} . The prediction is shown relative to the NLO one.
The NNLO correction is negative across the full kinematic range and, overall,
behaves very similarly to the one for the rapidity-inclusive ratio
$R_{3/2}(H_T)$. This remains the case as $y^*$ increases, at least in the range
of rapidities considered here.

\section{Conclusions}\label{sec:conclusion}

In this work we present for the first time NNLO-accurate predictions for
three-jet rates at the LHC.  We compute differential distributions for typical
jet observables like $H_T$ and the transverse momentum of the $i$th leading
jet, $i=1,2,3$, as well as differential three-to-two jet ratios. Scale
dependence is the main source of  theoretical uncertainty for this process at
NLO, and it gets significantly reduced after the inclusion of the NNLO QCD
corrections.  Notably, the three-to-two jet ratios stabilize once the
second-order QCD corrections are accounted for.

A central goal of the present work is to demonstrate the feasibility of
three-jet hadron collider computations with NNLO precision. With this
proof-of-principle goal attained, one can now turn one's attention to the broad
landscape of phenomenological applications for three-jet production at the LHC.
Examples include studies of event-shapes
\cite{Banfi:2004nk,Banfi:2010xy,Aad:2020fch}, determination of the running of
the strong coupling constant $\alpha_s$ through TeV scales and resolving the
question of scale setting in multi-jet production. Another major benefit from
having NNLO--accurate predictions is the reliability of the theory uncertainty
estimates.

On the technical side, the enormous computational cost of the present
calculation ($\sim 10^6$ CPUh) makes it clear that further refinements in the
handling of real radiation contributions to NNLO calculations are desirable.

\newpage
\section{Note added: erratum to the published version}

In the original publication, we evaluated the two-loop finite remainder
function ${\cal R}^{(2) l.c.}(s_{12})$ defined in equation (2) with an
incorrect colour factor.  This oversight was due to a missing conversion
factor between the conventions for the colour generator $T^a_{ij}$ used by the
authors of ref.~\cite{Abreu:2021fuk} (see \cite{Abreu:2019odu}, section 2
before equation 2.3) and our convention (see ref.~\cite{Czakon:2014oma},
appendix A).  By convention, the generators in ref.~\cite{Abreu:2021fuk} are
normalised such that $\Tr T^a T^b = \delta^{ab}$.  In our convention we use
$\Tr T^a T^b = \frac{1}{2}\delta^{ab}$, which implies a factor of $\sqrt{2}$
per appearing colour generator $T^a_{ij}$.  The following table lists the
colour factors and the conversion coefficient for the square of a colour factor
as it appears in the squared matrix element for each partonic channel:
\begin{center}
\begin{tabular}{c|c|c}
Channel                   & Colour factor $\mathcal{C}$ &
   $(|\mathcal{C}|^2)_{\text{our}}/(|\mathcal{C}|^2)_{\text{ref. [1]}}$  \\
\hline
$0 \to ggggg$             & $\Tr T^aT^bT^cT^dT^e$       & 64 \\
$0 \to gggq\bar{q}$       & $ (T^aT^bT^c)_{ij} $        & 8  \\
$0 \to gQ\bar{Q}q\bar{q}$ & $ (T^a)_{ij} \delta_{kl}$   & 2 
\end{tabular}
\end{center}
These conversion factors should have been included in our original calculation,
and we include them now in this erratum. These factors are sizable and have
implications on the phenomenology. In this version of the document, we provide
the corrected plots of the original publication.  The NNLO prediction increases
flatly by about $\approx 10\%$.  This implies that the double virtual
contribution is about $\approx 10\%$ of the total NNLO cross-section in
contrast to our previous findings of $\approx 2\%$.  With this, the naive
estimate for corrections from sub-leading colour terms would correspond to
$~1\%$ corrections of the NNLO QCD prediction.

\begin{acknowledgments}
We would like to thank Manuel Alvarez, Javier Llorente and Jennifer Roloff for
helpful discussions about ATLAS jet measurements. The work of M.C. was
supported by the Deutsche Forschungsgemeinschaft under grant 396021762 - TRR
257. The research of A.M. and R.P. has received funding from the European
Research Council (ERC) under the European Union's Horizon 2020 Research and
Innovation Programme (grant agreement no. 683211). A.M. was also supported by
the UK STFC grants ST/L002760/1 and ST/K004883/1. A.M.  acknowledges the use of
the DiRAC Cumulus HPC facility under Grant No. PPSP226.  Simulations were
performed with computing resources granted by RWTH Aachen University under
project rwth0414.
\end{acknowledgments}

\end{document}